\documentclass[aps,prl,floatfix,superscriptaddress,twocolumn,showpacs]{revtex4}

\usepackage{dcolumn}
\usepackage{graphicx}
\usepackage{epsf}
\usepackage{amsmath}
\usepackage{bm}
\usepackage{times}
\usepackage{nicefrac}

\newcolumntype{.}{D{x}{}{-1}}

\def\half{{\textstyle{\frac12}}}
\def\LZa{{L(Z\alpha)}}
\def\LZasquared{{L^2(Z\alpha)}}

\begin{document}

\title{Calculation of the One-- and Two--Loop Lamb Shift
       for Arbitrary Excited Hydrogenic States}

\author{Andrzej Czarnecki}
\affiliation{Department of Physics, University of Alberta,
  Edmonton, AB, Canada T6G 2J1}

\author{Ulrich D. Jentschura}
\affiliation{Max--Planck--Institut f\"ur Kernphysik,
Saupfercheckweg 1, 69117 Heidelberg, Germany}

\author{Krzysztof Pachucki}
\affiliation{Institute of Theoretical Physics,
Warsaw University, ul.~Ho\.{z}a 69, 00--681 Warsaw, Poland}

\begin{abstract}
General expressions for quantum electrodynamic corrections 
to the one--loop self-energy [of order $\alpha\,(Z\alpha)^6$] and  
for the two-loop Lamb shift [of order $\alpha^2\,(Z\alpha)^6$]
are derived. The latter includes all diagrams with closed fermion 
loops. The general results are valid for arbitrary excited non-$S$
hydrogenic states and for the normalized Lamb shift difference of $S$ states,
defined as $\Delta_n = n^3 \,\Delta E(nS) - \Delta E(1S)$.
We present numerical results for one-loop and 
two-loop corrections for excited $S$, $P$ and $D$ states.
In particular, the normalized Lamb shift difference of $S$ states
is calculated with an uncertainty of order 0.1~kHz.
\end{abstract}

\pacs{12.20.Ds, 31.30.Jv, 31.15.-p, 06.20.Jr}

\maketitle

The theory of quantum electrodynamics, when applied to the hydrogen atom
and combined with accurate measurements~\cite{FiEtAl2004etal,BeEtAl1997etal},
leads to the most accurately determined physical constants
today~\cite{MoTa2005} and to accurate
predictions for transition frequencies.
Of crucial importance are higher-order corrections to the bound-state
energies, which involve both purely relativistic 
atomic-physics effects and are mixed with the 
quantum electrodynamic (QED) corrections.
In general, this leads to a double expansion for the 
energy shifts, both in terms of the QED coupling $\alpha$ (the fine-structure
constant) and the nuclear charge number $Z$.

As is well known, the leading one-loop energy shifts 
(due to self-energy and vacuum polarization)
in hydrogenlike systems are of order $\alpha\,(Z\,\alpha)^4$ in 
units of the electron mass. 
Analytic calculations for higher excited states in the order
$\alpha\,(Z\,\alpha)^6$ are extremely demanding.
For non-$S$ states, the $\alpha\,(Z\,\alpha)^6$ corrections
have been obtained recently~\cite{JeEtAl2003}. 
However, excited $S$ states are very important for spectroscopy, 
and the corresponding gap in our knowledge is filled in the current Letter 
(see Table~\ref{table1}).
Regarding the two-loop correction, complete results
for the $\alpha^2\,(Z\,\alpha)^4$ effect were
obtained in 1970 (see Ref.~\cite{TwoLoop1970}).
Here, we derive general expressions which allow the determination of the 
entire two-loop $\alpha^2\,(Z\,\alpha)^6$ correction, 
for all non-$S$ hydrogenic states and the 
normalized difference 
$\Delta_n \equiv n^3 \, \Delta E(nS) - \Delta E(1S)$,
including the nonlogarithmic term.  
Together with other available analytic~\cite{Pa2001,PaJe2003} and 
numerical calculations for the $1S$ state~\cite{YeInSh2005}, 
our results allow for a much improved understanding of the 
higher-order two-loop corrections for a general excited hydrogenic 
states, and pave the way for an improved determination 
of fundamental constants from hydrogen spectroscopy.

The one-loop
bound-state self-energy, for the states under investigation here,
can be written as
\begin{align}
& \delta^{(1)}E = \frac{\alpha (Z\alpha)^4}{\pi n^3} 
\left\{ A_{40} 
+ (Z \alpha)^2 
\left[ A_{61}\ln[(Z \alpha)^{-2}] + A_{60} \right]\right\},
\nonumber
\end{align}
where the indices of the coefficients indicate the 
power of $Z\,\alpha$ and the power of the logarithm,
respectively. 
We work in $D = 4 - 2 \,\epsilon$ spacetime dimensions, and the 
dimension of space is $d = 3 - 2 \,\epsilon$.
Units are chosen so that $\hbar = c = \epsilon_0 = 1$, and the 
electron mass is unity.
A nonrelativistic, ``Bethe-style''~\cite{Be1947} calculation of the 
contribution due to ultrasoft photons, in the dipole approximation,
leads to a dimensionally regularized energy shift $E_{L0}$,
\begin{align}
\label{EL0}
E_{L0} = &
- \frac{4\alpha}{3\pi} \, \frac{(Z\,\alpha)^4}{n^3} \, \ln k_0 
\\
& \quad + Z\,\alpha^2
\left\{ \frac{2}{3\varepsilon} + \frac{10}{9} + \frac43\ln[(Z\,\alpha)^{-2}] 
\right\} \langle \delta^d(r) \rangle \,,\nonumber
\end{align}
where $\ln k_0 = \frac{n^3}{2(Z\alpha)^4} \left< p^i 
\left( H - E \right) \ln \left[ 2 | H - E|/(Z\alpha)^2 \right]
p^i \right>$ is the Bethe logarithm, 
and $\delta^d(r) = \vec{\nabla}^2 V/(4 \pi)$ 
is a $d$-dimensional Dirac delta function
obtained via the action of the Laplacian on the 
$d$-dimensional Coulomb potential
$V(r) = -Z\,\alpha \, r^{2-d}\, \left[
\Gamma\left(\frac{d}{2} - 1\right)\,\pi^{1-d/2} \right]$.
All matrix elements $\langle \cdot \rangle$ are to be evaluated with regard
to the reference state, as given by a nonrelativistic
(Schr\"{o}dinger--Pauli) wave function, and 
the summation convention is used throughout this Letter.

\begin{table}[htb]
\begin{minipage}{8.6cm}
\caption{\label{table1} Values of the nonlogarithmic 
self-energy correction $A_{60}$
(``relativistic Bethe logarithm'') for higher excited $S$ states.}
\begin{tabular}{r.@{\hspace{0.3in}}r.}
\hline
\hline
$n$ & \multicolumn{1}{c}{$A_{60}(nS)$} &
$n$ & \multicolumn{1}{c}{$A_{60}(nS)$} \\
\hline
1 & -30x.924\,149\,46(1) & 5 & -31x.455\,393(1)\\
2 & -31x.840\,465\,09(1) & 6 & -31x.375\,130(1)\\
3 & -31x.702\,501(1) & 7 & -31x.313\,224(1) \\
4 & -31x.561\,922(1) & 8 & -31x.264\,257(1) \\
\hline
\hline
\end{tabular}
\end{minipage}
\end{table}

Following~\cite{Pa1993,JePa1996,JeEtAl2003}, we now consider 
corrections due to the 
relativistic Hamiltonian, the quadrupole term and the relativistic 
and retardation corrections to the current. 
The relativistic correction to the Hamiltonian is
\begin{equation}
\label{defHR}
H_R = - \frac{\vec p^{\,4}}{8} +
\frac{\pi}{2}\,Z\,\alpha\,\delta^d(r)+\frac{1}{4}\,
\sigma^{ij}\,\nabla^i V\,p^j\,.
\end{equation}
Here, $\sigma^{ij} \equiv \frac{1}{2\,{\rm i}}\,[\sigma^i,\,\sigma^j]$.
The resulting, dimensionally regularized, correction to the Bethe logarithm is 
\begin{align}
\label{EL1intermediate}
& E_{L1} =
\frac{\alpha}{\pi}\,\frac{(Z\,\alpha)^6}{n^3}\,\beta_1 +
\frac{\alpha}{3\,\pi}\,
\left\{\frac{1}{2\,\varepsilon}+\frac{5}{6}+
\LZa \right\} \nonumber \\ 
& \times \left< \frac{1}{8} \vec{\nabla}^4 V +
\frac{\rm i}{4} \sigma^{ij} p^i
\vec{\nabla}^2 V\,p^j + 2 H_R\,{\overline G} \vec{\nabla}^2 V \right>,
\end{align}
where $L(Z\alpha) \equiv \ln\left[\half (Z\alpha)^{-2}\right]$, 
and ${\overline G} = 1/(E-H)'$ is the reduced Green function;
$\beta_1$ is a generalized Bethe logarithm,
\begin{align}
\label{defbeta1}
& \frac{(Z\alpha)^6}{n^3}\beta_1 
= -\frac{4}{3} \left<  H_R \, {\overline G} \, p^i 
(H-E) \ln\left[\frac{|H-E|}{(Z\,\alpha)^2}\right] p^i \right>
\nonumber\\
& +\frac{2}{3} \,  \sum_{n,m}
\frac{\langle \phi | p^i | n \rangle
\langle n | H_R | m \rangle
\langle m | p^i | \phi \rangle}{E_m-E_n} \biggl\{ (E_n-E)
\nonumber\\
& \quad \times \ln\left[\frac{|E_n - E|}{(Z\alpha)^2}\right]
- (E_m-E) \ln\left[\frac{|E_m - E|}{(Z\alpha)^2}\right] \biggr\}
\nonumber\\
& +\frac{2}{3} \, \left< H_R \right> \,
\left< p^i \left\{ 1+\ln\left[\frac{|H-E|}{(Z\,\alpha)^2}\right] \right\}
p^i \right>\,.
\end{align}
We temporarily restore the reference state $\phi$ in the 
notation of the matrix element, and the sums over $n$ and $m$
include both the discrete as well as the continuous part of the spectrum. 
The argument of the logarithm in
$\beta_1$ is $\ln[|H-E|/(Z\,\alpha)^2]$,
not $\ln[2 |H-E|/(Z\,\alpha)^2]$ as in $\ln k_0$, and this fact is 
important for the precise definition of $\beta_1$,
and of all other generalized Bethe logarithms in the following.

In the dimensional scheme,
the quadrupole correction $E_{L2}$, which was denoted 
as $F_{\rm nq}$ in former work~\cite{Pa1993,JePa1996}, is found to be
expressible as $E_{L2} = {\cal D}_2 + {\cal F}_2$, where
\begin{align}
& {\cal D}_2 =
\frac{\alpha}{\pi} \left<
\frac{2 (\vec{\nabla} V)^2}{3} \right>\,
\left[ \frac{1}{\varepsilon}+\frac{103}{60}+ 2 L(Z\alpha) \right] 
\nonumber\\
& + \left< \frac{\vec{\nabla}^4 V}{40} \right>\,
\left[\frac{1}{\varepsilon}+\frac{12}{5}
+2 L(Z\alpha)\right]
\nonumber\\
& + \left< \frac{\vec{\nabla}^2V\,\vec p^{\,2}}{6} \right>\,
\left[\frac{1}{\varepsilon}+\frac{34}{15} +2 L(Z\alpha)\right]\,,
\nonumber
\end{align}
and ${\cal F}_2$ contains the generalized Bethe logarithm
$\beta_2$,
\begin{align}
& {\cal F}_2 = \frac{\alpha (Z\alpha)^6 \beta_2}{\pi n^3} = 
\frac{\alpha}{\pi} \int \frac{d\Omega_{\vec n}}{4\pi} 
\left( \delta^{ij} - n^i\,n^j\right)\,
\nonumber\\
& \times \left\{
\left< p^i (\vec n\cdot\vec r)^2 
(H-E)^3 \ln\left[\frac{|H-E|}{(Z\,\alpha)^2}\right] \, p^j \right>
\right.
\nonumber\\
& \left. - \left< p^i(\vec n\cdot\vec r) 
(H-E)^3 \ln\left[\frac{|H-E|}{(Z\,\alpha)^2}\right]\,
p^j(\vec n\cdot\vec r)\,\right> \right\}\,.
\nonumber
\end{align}
Here, $\vec n$ is a three-dimensional unit vector, and we 
integrate over the entire solid angle $\Omega_{\vec n}$.
Throughout this Letter, $\vec{\nabla}^2$ and 
$\vec{\nabla}^4$ are understood to exclusively act on the 
quantity immediately following the operator, 
i.e.~$\left< \vec{\nabla}^2V\,\vec p^{\,2}\right> = 
\left< (\vec{\nabla}^2V)\,\vec p^{\,2}\right>$,
$\left< \vec{\nabla}^2 V\,{\overline G} \,H_R \right> =
\left< (\vec{\nabla}^2 V)\,{\overline G} \,H_R \right>$ etc.

The correction $E_{L3}$ to the transition current
reads $E_{L3} = {\cal D}_3 + {\cal F}_3$,
where ${\cal F}_3 = \alpha (Z\alpha)^6 \beta_3/\pi n^3$ contains
the generalized Bethe logarithm $\beta_3$, and 
\begin{align}
& {\cal D}_3 =
-\frac{\alpha}{\pi}
\left[\frac{2}{3\varepsilon}+
\frac{10}{9}+
\frac43 \LZa{} \right]
\left< \frac{\vec{\nabla}^2V\vec p^{\,2}}{4} +
\frac{\bigl(\vec{\nabla} V\bigr)^2}{2} \right> ,
\nonumber\\
& {\cal F}_3 =
\frac{2\alpha}{3\pi}
\left< j^i\, (H-E) 
\ln\left[\frac{|H-E|}{(Z\,\alpha)^2}\right]
p^i \right>.
\nonumber
\end{align}
Here, $j^i = p^i \vec{p}^{\,2}+\frac12\,\sigma^{ij}\nabla^j V$,
and $\nabla^i \equiv \partial/\partial r^i$ denotes the 
derivative with respect to the $i$th Cartesian coordinate.
The divergences (in $\varepsilon$) in the corrections to the 
Bethe logarithm are compensated by high-energy virtual 
photons, which in nonrelativistic QED (NRQED) are given by effective 
operators. From a generalized Dirac equation (see Chap.~7 of
Ref.~\cite{ItZu1980}), one easily obtains the effective
one-loop potential
\begin{equation}
\delta^{\rm (1)} V =
-\frac{1}{6\,\varepsilon}\,\frac{\alpha}{\pi}\,
\vec{\nabla}^2 V
+\frac{\alpha}{4\,\pi}\,\sigma^{ij}\nabla^i V\,p^j,
\end{equation}
which in leading order gives rise to the correction 
$\left< \delta^{\rm (1)} V \right>$. This correction
is a contribution to the middle-energy part $E_{M}$,
which originates from 
from high-energy virtual photons, with electron momenta 
of order $Z\alpha$. The corrections of relative order $(Z\,\alpha)^2$ to 
$\left< \delta^{\rm (1)} V \right>$ involve relativistic
corrections to the wave function and to the operators,
and a two-Coulomb-vertex scattering amplitude. 
The sum is 
\begin{align}
\label{defEM1EM2}
& E_{M} = \left< \delta^{\rm (1)} V \right> + 
2\,\left\langle \delta^{\rm (1)}V \,
{\overline G}\,H_R\right\rangle
\\
& +\frac{\alpha}{\pi}\,\left(\frac{1}{192} - \frac{1}{48 \varepsilon}\right)
\langle\vec{\nabla}^4V+2\,{\rm i}\,
\sigma^{ij}\,p^i\vec{\nabla}^2 V\, p^j\rangle
\nonumber\\
& -\frac{\alpha}{32\pi}\,\left\langle
\left\{\vec p^{\,2},\vec{\nabla}^2V
+2\,\sigma^{ij}\,\nabla^i V\,p^j \right\}\right\rangle
\nonumber\\
& - \frac{\alpha}{\pi}\,\left(\frac{11}{240} + \frac{1}{40\varepsilon}\right)\,
\langle\vec{\nabla}^4V \rangle
+ \frac{\alpha}{\pi}\,\left(\frac{11}{48} - \frac{1}{3\,\varepsilon} \right)
\,\left< (\vec{\nabla} V)^2 \right>\,.
\nonumber
\end{align}
The complete one-loop result 
$\delta^{\rm (1)} E = 
E_{L0} + E_{L1} + E_{L2} + E_{L3} + E_{M}$ 
reads
\begin{widetext}
\begin{align}
\label{delta1LE}
& \delta^{\rm (1)} E = 
\frac{\alpha}{\pi}\,\frac{(Z\,\alpha)^4}{n^3} \,
\left( 
\left[ \frac{10}{9}+\frac{4}{3}\,\ln\left[(Z\,\alpha)^{-2}\right]\right]\,
\delta_{l0} - \frac43\,\ln k_0 \right)
+\frac{\alpha}{4\,\pi}\,
\left< \sigma^{ij}\nabla^i V p^j \right>
+ \frac{\alpha}{\pi}\, \frac{(Z\,\alpha)^6}{n^3}\, 
\left( \beta_1+\beta_2+\beta_3 \right)
\nonumber \\ 
& + \frac{\alpha}{\pi}\, \left\{
\left(\frac{5}{9}+ \frac{2}{3}\, \LZa \right)\,
\left< \vec{\nabla}^2 V\,{\overline G} \,H_R \right>
+\frac{1}{2}\,\left< \sigma^{ij}\nabla^i
V p^j\,{\overline G} \,H_R \right>
+\left(\frac{779}{14400}+\frac{11}{120}\,
\LZa{} \right)\,\left< \vec{\nabla}^4 V\right> 
\right.
\nonumber \\ 
& \left.
+\left(\frac{23}{576}+\frac{1}{24} \LZa\right)
\,\langle 2\,{\rm i}\,\sigma^{ij}p^i\vec{\nabla}^2V p^j \rangle
+\left(\frac{589}{720}+\frac{2}{3} \LZa \right)
\langle(\vec{\nabla} V)^2\rangle
+\frac{3}{80}\,\left< \vec p^{\,2} \vec{\nabla}^2 V \right>
-\frac{1}{8}\,\left< \vec{p}^{\,2} \sigma^{ij}\nabla^i V p^j \right> \right\}.
\end{align}
\end{widetext}
The matrix elements in this result can be 
evaluated using standard techniques.
In terms of the notation of Ref.~\cite{JeEtAl2003},
we have ${\cal L} = \sum_{i=1}^3 \beta_i$.
Our general result (\ref{delta1LE}), evaluated for hydrogenic 
states, reproduces the known logarithmic term $A_{61}$,
and is consistent with all formulas reported for the nonlogarithmic term in 
Eqs.~(10) and (12) of Ref.~\cite{JeEtAl2003}. 
The evaluation of ${\cal L}$ is a demanding 
numerical calculation, and numerical values for non-$S$ states
have been presented in Table I of~Ref.~\cite{JeEtAl2003}. Taking advantage 
of the result~\cite{Pa1993} for $1S$ and the 
validity of Eq.~(\ref{delta1LE}) for the $nS$-$1S$ difference,
we can now proceed to indicate results for the nonlogarithmic 
term $A_{60}$ for $nS$ states, an evaluation made possible 
by our generalized NRQED approach (see Table~\ref{table1}).

A generalization of our NRQED approach leads to the following
general result for the $\alpha^2 \, (Z\alpha)^6$-term 
of the complete 
two-loop Lamb shift (including all diagrams with closed fermion 
loops), 
\begin{widetext}
\begin{align}
\label{delta2LE}
\delta^{(2)} E =& 
\frac{\alpha^2 (Z\alpha)^6}{\pi^2 n^3} \left\{ B_{62}\ln^2[(Z \alpha)^{-2}] 
+ B_{61}\ln[(Z \alpha)^{-2}] + B_{60} \right\}
= 
\frac{\alpha^2 (Z\alpha)^6}{\pi^2 n^3} 
\left\{ b_L + \beta_{4} + \beta_{5} + 
\left[\frac{38}{45} + 
\frac{4}{3} \LZa{} \right]\, N \right\}
\nonumber \\ 
& 
+ \left(\frac{\alpha}{\pi}\right)^2\,
\left[-\frac{42923}{259200} 
+ \frac{9}{16} \,\zeta(2)\,\ln (2)
- \frac{5}{36} \,\zeta(2)
- \frac{9}{64}\,\zeta(3)
+ \frac{19}{135} \LZa
+ \frac{1}{9} \LZasquared \right]\,
\left< \vec{\nabla}^2 V\,{\overline G} \, \vec{\nabla}^2 V \right> 
\nonumber \\ 
& 
+ \left(\frac{\alpha}{\pi}\right)^2\,
\left[ \frac{2179}{10368} 
- \frac{9}{16} \,\zeta(2)\,\ln (2)
+ \frac{5}{36} \,\zeta(2)
+ \frac{9}{64}\,\zeta(3) \right]\,
\left< \vec{\nabla}^2 V\,{\overline G} \, \vec{p}^{\,4} \right> 
\nonumber \\ 
& 
+ \left(\frac{\alpha}{\pi}\right)^2\,
\left[ -\frac{197}{1152} 
+ \frac{3}{8} \,\zeta(2)\,\ln (2)
- \frac{1}{16} \,\zeta(2)
- \frac{3}{32}\,\zeta(3) \right]\,
\left< \vec{p}^{\,4} \,{\overline G} \, \sigma^{ij}\,\nabla^i V\,p^j \right> 
\nonumber \\ 
& 
+ \left(\frac{\alpha}{\pi}\right)^2\,
\left[ \frac{233}{576} 
- \frac{3}{4} \,\zeta(2)\,\ln (2)
+ \frac{1}{8} \,\zeta(2)
+ \frac{3}{16}\,\zeta(3) \right]\,
\left< \sigma^{ij}\,\nabla^i V\,p^j \,{\overline G} \, 
\sigma^{ij}\,\nabla^i V\,p^j \right> 
\nonumber \\ 
& +\left(\frac{\alpha}{\pi}\right)^2\,
\left[-\frac{197}{2304} 
+ \frac{3}{16} \,\zeta(2)\,\ln (2)
- \frac{1}{32} \,\zeta(2)
- \frac{3}{64}\,\zeta(3)\right]\,
\left< 
\left\{ \vec{p}^{\,2} , \vec{\nabla}^2 V + 2\,\sigma^{ij}\,\nabla^i V\,p^j\right\}
\right> 
\nonumber \\ 
& +\left(\frac{\alpha}{\pi}\right)^2\,
\left[-\frac{83}{1152}  
+ \frac{17}{8} \zeta(2)\,{\ln(2)}
- \frac{59}{72}\,\zeta(2) 
- \frac{17}{32}\,\zeta(3) \right]\,
\left< \left(\vec{\nabla} V \right)^2\right>
\nonumber \\ 
& +\left(\frac{\alpha}{\pi}\right)^2\,
\left[-\frac{87697}{345600} 
+ \frac{9}{10}\zeta(2)\,\ln (2)
- \frac{2167}{9600}\zeta(2)
- \frac{9}{40}\zeta(3)
+ \frac{19}{270} \LZa
+ \frac{1}{18} \LZasquared \right]\,
\left< \vec{\nabla}^4 V \right>
\nonumber \\ 
& + \left(\frac{\alpha}{\pi}\right)^2\,
\left[-\frac{16841}{207360}  
- \frac{1}{5}\, \zeta(2)\,\ln(2)
+ \frac{223}{2880} \, \zeta(2)
+ \frac{1}{20}\,\zeta(3)
+ \frac{1}{24}\, \LZa
\right]\,
\left< 2\,{\rm i}\,\sigma^{ij}\, p^i\,\vec{\nabla}^2V\,p^j\right>\,.
\end{align}
\end{widetext}
The leading $\alpha^2 (Z\alpha)^4$-term, given by the  
$B_{40}$ coefficient, is well known 
and therefore not included here (for a review 
see e.g.~Appendix A of~Ref.~\cite{MoTa2005}). The above expression is valid for 
$P$, $D$ states, and for the normalized difference $\Delta_n$ of $S$ states.
The quantity $N$ is defined in terms of 
the notation adopted in Refs.~\cite{Pa2001,Je2003jpa},
and the two-loop Bethe logarithm $b_L$ is defined in 
Refs.~\cite{PaJe2003,Je2004b60}. Although $b_L$  
has been determined numerically only for $S$ states (see Ref.~\cite{Je2004b60}), 
it represents a well-defined quantity
for all hydrogenic states. 
The logarithmic sum $\beta_{4}$ 
is given by Eq.~(\ref{defbeta1}),
with the replacement $H_R
\to \frac14\,\sigma^{ij}\nabla^i\,V\,p^j$.
Finally, we have 
\begin{equation}
\frac{(Z\alpha)^6}{n^3}\beta_5 = \frac12\,\left< \sigma^{ij}\nabla^j\,
(H-E)\,\ln\left[\frac{|H-E|}{(Z\alpha)^2} \right] \, p^i\right>.
\end{equation}

Evaluating the general expression (\ref{delta2LE}) for $P$ states,
we confirm that $B_{62}(nP) = \frac{4}{27} \frac{n^2 - 1}{n^2}$.  
Furthermore, we obtain the 
results
\begin{align}
B_{61}(nP_{1/2}) =& \frac43\, N(nP) + 
\frac{n^2 - 1}{n^2}
\left(\frac{166}{405} -\frac{8\ln 2}{27} \right)\,,
\nonumber\\
B_{61}(nP_{3/2}) =& \frac43\, N(nP) + \frac{n^2 - 1}{n^2}
\left(\frac{31}{405} -\frac{8\ln 2}{27} \right)\,.
\end{align}
Numerical values for $N(nP)$ can be found in Eq.~(17) of~\cite{Je2003jpa}.
Regarding the
nonlogarithmic term $B_{60}$, we fully confirm results for the
fine-structure difference of $P$ states~\cite{JePa2002}.
A further important conclusion to 
be drawn from Eq.~(\ref{delta2LE}) is that 
all logarithmic two-loop terms
of order $\alpha^2\,(Z\alpha)^6$ vanish for states with
orbital angular momentum $l \geq 2$.

We have also verified that the two-loop result 
(\ref{delta2LE}) is consistent
with the normalized $S$-state difference $\Delta_n$ for the 
logarithmic terms $B_{62}$ and $B_{61}$,
as derived in Ref.~\cite{Pa2001} (using a completely
different method). Evaluating all matrix elements
in Eq.~(\ref{delta2LE}), we are now in the position
to obtain the $n$-dependence of the nonlogarithmic term, which we write
as $B_{60}(nS) - B_{60}(1S) = b_L(nS) - b_L(1S) + A(n)$, 
where $A(n)$ is the additional contribution
beyond the $n$-dependence of the two-loop Bethe logarithm. 
The result for $A(n)$ is
\begin{widetext}
\begin{align}
\label{An}
& A(n) =
\left( \frac{38}{45} - \frac43\,\ln(2) \right) \, [N(nS) - N(1S)]  
- \frac{337043}{129600} - \frac{94261}{21600\,n} + \frac{902609}{129600\,n^2} 
+ \left( \frac{4}{3}  
- \frac{16}{9\,n}
+ \frac{4}{9\,n^2} \right) \, \ln^2 (2)
\\
&
+ \left( -\frac{76}{45} 
+ \frac{304}{135\,n}
- \frac{76}{135\,n^2} \right) \,\ln(2) 
+ \left( - \frac{53}{15}
+ \frac{35}{2\,n}
- \frac{419}{30\,n^2} \right) \, \zeta(2)\,\ln(2) 
+ \left( \frac{28003}{10800} 
- \frac{11}{2\,n} 
+ \frac{31397}{10800\,n^2} \right) \, \zeta(2) 
\nonumber\\
& + \left( \frac{53}{60} 
- \frac{35}{8 n}
+ \frac{419}{120 n^2} \right) \,\zeta(3)  
+ \left( \frac{37793}{10800} 
+ \frac{16}{9}\ln^2(2)
- \frac{304}{135}\ln (2) 
+ 8\zeta(2)\ln(2) 
- \frac{13}{3}\zeta(2) 
- 2\zeta(3)\right) 
\left[ \gamma + \Psi(n) - \ln(n) \right] \,.
\nonumber
\end{align} 
\end{widetext}
Numerically, $A(n)$ is found to be much smaller than 
$b_L(nS) - b_L(1S)$, which implies that the main
contribution to $B_{60}(nS) - B_{60}(1S)$ 
is exclusively due to the two-loop Bethe logarithm.
As an example, we consider 
$A(5) = 0.370~042$ and $B_{60}(5S) - B_{60}(1S) = 21.2(1.1)$,
where the error is due to the numerical uncertainty of the 
two-loop Bethe logarithm $b_L(5S)$ (see Ref.~\cite{Je2004b60}).

\begin{table}[htb]
\begin{minipage}{8.6cm}
\caption{\label{table2} Theoretical 
values of the normalized 
Lamb-shift difference $\Delta_n = n^3 \,\Delta E(nS) - \Delta E(1S)$,
based on the results reported in this Letter [see Eq.~(\ref{An})].
Units are kHz.}
\begin{tabular}{r@{\hspace*{0.5cm}}.r@{\hspace*{0.5cm}}.}
\hline
\hline
$n$ & \multicolumn{1}{c}{$\Delta_n$} &
$n$ & \multicolumn{1}{c}{$\Delta_n$} \\
\hline
2 &    187225x.70(5) & 12 &    279988x.60(10) \\
3 &    235070x.90(7) & 13 &    280529x.77(10) \\
4 &    254419x.32(8) & 14 &    280962x.77(10) \\
5 &    264154x.03(9) & 15 &    281314x.61(10) \\
6 &    269738x.49(9) & 16 &    281604x.34(11) \\
7 &    273237x.83(9) & 17 &    281845x.77(11) \\
8 &    275574x.90(10) & 18 &    282049x.05(11) \\
9 &    277212x.89(10) & 19 &    282221x.81(11) \\
10 &    278405x.21(10) & 20 &    282369x.85(11) \\
11 &    279300x.01(10) & 21 &    282497x.67(11) \\
\hline
\hline
\end{tabular}
\end{minipage}
\end{table}

The test of standard model theories and the 
determination of fundamental constants (specifically, of the Rydberg 
constant and of the electron mass) 
provide the main motivations
for carrying out the QED calculations in ever higher orders
of approximation. Recently, our knowledge of the ground-state Lamb shift 
has been improved by a fully numerical calculation of the 
two-loop self-energy~\cite{YeInSh2005}. However, because of the 
structure of the hydrogen spectrum, the decisive quantity
for the determination of the Rydberg constant from spectroscopic 
data is the normalized difference $\Delta_n$ of the 
$nS-1S$ Lamb-shift. Elucidating discussions regarding the latter
point can be found near Eqs.~(2) and (3) of Ref.~\cite{UdEtAl1997},
and in Appendix A of Ref.~\cite{MoTa2005}.
Accurate theoretical values for $\Delta_n$ can be inferred
from the results reported here and are compiled in Table~\ref{table2}.
The Rydberg constant is currently known to a relative accuracy of 
$6.6 \times 10^{-12}$, limited essentially by the 
experimental accuracy of the $2S-8D$ and $2S-12D$ measurements 
(see Table~V of~\cite{MoTa2005}). Using the improved theory as 
presented in this Letter, it will become possible to determine 
the Rydberg constant to an accuracy on the level of $10^{-14}$,
provided the ongoing experiments concerning the 
hydrogen $1S-3S$ transition~\cite{UdPriv2005,ArPriv2004}
reach a sub-kHz level of accuracy. 

The authors acknowledge helpful conversations with 
P. J. Mohr and R. Bonciani. 
This work was supported by EU grant No.~HPRI-CT-2001-50034. 
A.C. acknowledges support by the 
Natural Sciences and Engineering Research Council of Canada.
U.D.J. acknowledges support from the Deutsche Forschungsgemeinschaft
(Heisenberg program).


\begin{thebibliography}{10}

\bibitem{FiEtAl2004etal}
{M.~Fischer~{\em et al.}}, Phys. Rev. Lett. {\bf 92},  230802  (2004).

\bibitem{BeEtAl1997etal}
{B.~de~Beauvoir {\em et al.}}, Phys. Rev. Lett. {\bf 78},  440  (1997).

\bibitem{MoTa2005}
P.~J. Mohr and B.~N. Taylor, Rev. Mod. Phys. {\bf 77},  1  (2005).

\bibitem{JeEtAl2003}
U.~D. Jentschura, E.-O. Le~Bigot, P.~J. Mohr, P. Indelicato, and G. Soff, Phys.
  Rev. Lett. {\bf 90},  163001  (2003).

\bibitem{TwoLoop1970}
T.~Appelquist and S.~J.~Brodsky, Phys. Rev. Lett. {\bf 24}, 562 (1970) and
  Phys. Rev. A {\bf 2}, 2293 (1970); R.~Barbieri, J.~A.~Mignaco and E.~Remiddi,
  Lett. Nuovo Cim. {\bf 3}, 588 (1970); B.~E.~Lautrup, A.~Peterman and E.~de
  Rafael, Phys. Lett. B {\bf 31}, 577 (1970).

\bibitem{Pa2001}
K. Pachucki, Phys. Rev. A {\bf 63},  042503  (2001).

\bibitem{PaJe2003}
K. Pachucki and U.~D. Jentschura, Phys. Rev. Lett. {\bf 91},  113005  (2003).

\bibitem{YeInSh2005}
V.~A. Yerokhin, P. Indelicato, and V.~M. Shabaev, Phys. Rev. A {\bf 71},
  R040101  (2005).

\bibitem{Be1947}
H.~A. Bethe, Phys. Rev. {\bf 72},  339  (1947).

\bibitem{Pa1993}
K. Pachucki, Ann. Phys. (N.Y.) {\bf 226},  1  (1993).

\bibitem{JePa1996}
U. Jentschura and K. Pachucki, Phys. Rev. A {\bf 54},  1853  (1996).

\bibitem{ItZu1980}
C. Itzykson and J.~B. Zuber, {\em Quantum Field Theory} (McGraw-Hill, New York,
  NY, 1980).

\bibitem{Je2003jpa}
U.~D. Jentschura, J. Phys. A {\bf 36},  L229  (2003).

\bibitem{Je2004b60}
U.~D. Jentschura, Phys. Rev. A {\bf 70},  052108  (2004).

\bibitem{JePa2002}
U.~D. Jentschura and K. Pachucki, J. Phys. A {\bf 35},  1927  (2002).

\bibitem{UdEtAl1997}
Th. Udem, A. Huber, B. Gross, J. Reichert, M. Prevedelli, M. Weitz, and T.~W.
  H\"{a}nsch, Phys. Rev. Lett. {\bf 79},  2646  (1997).

\bibitem{UdPriv2005}
Th. Udem, private communication (2005).

\bibitem{ArPriv2004}
O. Arnoult, private communication (2004).

\end{thebibliography}
\end{document}